\documentclass[12pt]{pnastwo}

\usepackage[dvips]{graphicx}
\usepackage{hyperref}

\hypersetup{
    colorlinks=true,
    linkcolor=blue,
    filecolor=blue,      
    urlcolor=blue,
}
 
\usepackage{amssymb,amsfonts,amsmath}

\copyrightyear{2018}
\issuedate{}
\volume{}
\issuenumber{}

\begin{document}
\tracingall


\title{Life Beyond the Solar System: Remotely Detectable Biosignatures}





\author{
Shawn Domagal-Goldman\affil{1}{NASA Goddard Space Flight Center}, Nancy Y. Kiang\affil{2}{NASA Goddard Institute for Space Studies}, Niki Parenteau\affil{3}{NASA Ames Research Center}, David C. Catling\affil{4}{Dept. Earth and Space Sciences / Astrobiology Program, University of Washington}, Shiladitya DasSarma\affil{5}{Institute of Marine and Environmental Technology, University of Maryland School of Medicine, Baltimore, Maryland}, Yuka Fujii\affil{6}{Earth-Life Science Institute, Tokyo Institute of Technology and NASA Goddard Institute for Space Studies}, Chester E. Harman\affil{7}{Columbia University and NASA Goddard Institute for Space Studies}, Adrian Lenardic\affil{8}{Rice University
}, Enric Pall\'{e}\affil{9}{Instituto de Astrof\'{i}sica de Canaria, Spain}, Christopher T. Reinhard\affil{10}{School of Earth and Atmospheric Sciences, Georgia Institute of Technology}, Edward W. Schwieterman\affil{11}{Dept. Earth
Sciences, University of California, Riverside, California}, Jean Schneider\affil{12}{Paris Observatory}, Harrison B. Smith\affil{13}{School of Earth and Space Exploration, Arizona State University}, Motohide Tamura\affil{14}{University of Tokyo / Astrobiology Center of NINS}, Daniel Angerhausen\affil{15}{Center for Space and Habitability, Bern University, Switzerland}, Giada Arney\affil{1}{}, Vladimir S. Airapetian\affil{16}{NASA Goddard Space Flight Center and American University}, Natalie M. Batalha\affil{3}{}, Charles S. Cockell\affil{17}{UK Centre for Astrobiology, School of Physics and Astronomy, University of Edinburgh}, Leroy Cronin\affil{18}{School of Chemistry, University of Glasgow, UK}, Russell Deitrick\affil{19}{Dept. Astronomy, University of Washington}, Anthony Del Genio\affil{2}{}, Theresa Fisher\affil{13}{}, Dawn M. Gelino\affil{20}{NASA Exoplanet Science Institute}, J. Lee Grenfell\affil{21}{Dept. Extrasolar Planets and Atmospheres, German Aerospace Centre}, Hilairy E. Hartnett\affil{13}{}, Siddharth Hegde\affil{22}{Carl Sagan Institute, and Cornell Center for Astrophysics and Planetary Science, Cornell University}, Yasunori Hori\affil{23}{
Astrobiology Center and National Astronomical Observatory of Japan}, Bet\"ul Ka\c{c}ar\affil{24}{Depts. of Molecular and Cellular Biology and Astronomy, University of Arizona}, Joshua Krissansen-Totten\affil{4}{}, Timothy Lyons\affil{11}{}, William B. Moore\affil{25}{Hampton University and National Institute of Aerospace}, Norio Narita\affil{26}{Dept. of Astronomy, The University of Tokyo}, Stephanie L. Olson\affil{11}{}Heike Rauer\affil{27}{German Aerospace Centre (DLR) Institute of Planetary Research
}, Tyler D. Robinson\affil{28}{Dept. Physics and Astronomy, Northern Arizona University}, Sarah Rugheimer\affil{29}{Centre for Exoplanets, School of Earth and Environmental Sciences, University of St. Andrews, UK}, Nick Siegler\affil{30}{Jet Propulsion Laboratory, California Institute of Technology}, Evgenya L. Shkolnik\affil{13}{}, Karl R. Stapelfeldt\affil{30}{}, Sara Walker\affil{31}{School of Earth and Space Exploration and Beyond Center for Fundamental Concepts in Science, Arizona State University}
}

\contributor{This white paper summarizes the products from the Exoplanet Biosignatures Workshop Without Walls (EBWWW). The following people also contributed to the EBWW. Science Organizing Committee (SOC): Daniel Apai, Shawn Domagal-Goldman, Yuka Fujii, Lee Grenfell, Nancy Y. Kiang, Adrian Lenardic, Nikole Lewis, Timothy Lyons, Hilairy Hartnett, Bill Moore, Enric Pall\'{e}, Niki Parenteau, Heike Rauer, Karl Stapelfeldt, Sara Walker. Online and In-Person Workshop Participants: SOC named above as well as Giada Arney, William Bains, Robert Blankenship, David Catling, Charles Cockell, David Crisp, Sebastian Danielache, Shiladitya DasSarma, Russell Deitrick, Anthony Del Genio, Drake Deming, Steve Desch, David Des Marais, Theresa Fisher, Sonny Harman, Erika Harnett, Siddharth Hegde, Yasunori Hori, Renyu Hu, Bet\"ul Ka\c{c}ar, Jeremy Leconte, Andrew Lincowski, Rodrigo Luger, Victoria Meadows, Adam Monroe, Norio Narita, Christopher Reinhard, Sarah Rugheimer, Andrew Rushby, Edward Schwieterman, Nick Siegler, Evgenya Skolnick, Harrison Smith, Motohide Tamura, Mike Tollion, Margaret Turnbull, and Mary Voytek.}

\maketitle
\keywords{exoplanets | habitability | biosignatures | astrobiology}
\clearpage
\setcounter{page}{1}
\begin{article}





\section{Introduction}
\begin{normalsize}
For the first time in human history, we will soon be able to apply the scientific method to the question ''Are We Alone?'' The rapid advance of exoplanet discovery, planetary systems science, and telescope technology will soon allow scientists to search for life beyond our Solar System through direct observation of extrasolar planets. This endeavor will occur alongside searches for habitable environments and signs of life within our Solar System. While these searches are thematically related and will inform each other, they will require separate observational techniques. The search for life on exoplanets holds potential through the great diversity of worlds to be explored beyond our Solar System. However, there are also unique challenges related to the relatively limited data this search will obtain on any individual world.
\end{normalsize}

This white paper reviews the scientific community's ability to use data from future telescopes to search for life on exoplanets. This material summarizes products from the Exoplanet Biosignatures Workshop Without Walls (EBWWW). The EBWWW was constituted by a series of online and in-person activities, with participation from the international exoplanet and astrobiology communities, to assess state of the science and future research needs for the remote detection of life on planets outside our Solar System. These activities culminated in five manuscripts, submitted for publication, which respectively cover: 1) a review of known and proposed biosignatures (Schwieterman et al., in press), 2) a review of O\textsubscript{2} as a biosignature as an end-to-end example of the contextual knowledge required to rigorously assess any claims of life on exoplanets (Meadows et al., in press); 3) a generalized statistical approach to place qualitative understanding and available data in a formal quantitative framework according to current understanding (Catling et al., in press); 4) identification of needs to advance that statistical framework, and to develop or incorporate other conceptual frameworks for biosignature assessment (Walker et al., in review), and 5) a review of the upcoming observatories - both planned and possible - that could provide the data needed to search for exoplanet biosignatures (Fujii et al., in review). These manuscripts were written by an interdisciplinary and international community of scientists, incorporating input from both an open public comment period and an anonymous journal peer review process. As such, they represent the community-wide scientific consensus on the state of the field, and on the research priorities to further the search for life on exoplanets.

\section{Progress Since 2015 Astrobiology Strategy}

\subsection{Expanding the library of signs of life}
Analyses of a planet's spectrum, even from a single spatial element, can yield information on the presence or absence of chemicals that absorb specific wavelengths of light. It is this limited information upon which many of our proposed biosignatures, as well as other features of the planet's environmental context, must be identified. Much of the history of remote detection of biosignatures focused on spectral features of specific biological byproducts or global phenomena resulting from life. A review of exoplanet biosignatures is presented in Schwieterman et al. (in press), updating a prior review by Des Marais et al. (2002), which was considered in the writing of the Astrobiology Strategy 2015 document. There have been three major developments in exoplanet biosignature science since 2015: the generation of a broader list of potential biosignatures, more comprehensively simulations of these signatures in the context of planetary environments, and consideration of abiotic means through which these signatures could be generated on both living and non-living worlds.

\begin{figure}[t]
\centering{\includegraphics[width=3.5in]{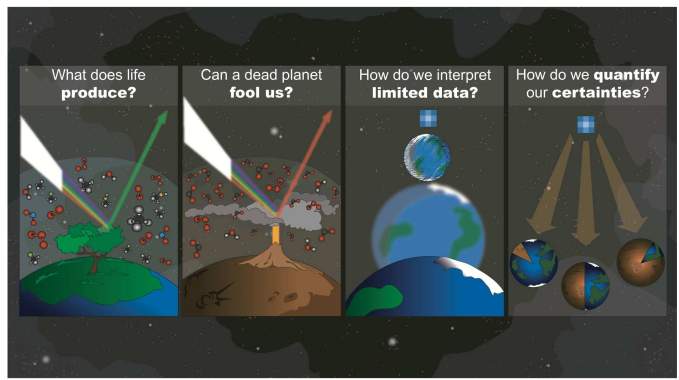}}
\caption{An overview of the past, present, and future of biosignature theory research. Research historically has focused on cataloguing lists of substances or physical features that yield spectral signatures as indicators of potential life on exoplanets. Recent progress has led to understanding of how non-living planets could produce similar signatures. In the future, the field should strive to utilize what are inherently limited data to deliver quantitative assessments of whether or not a given planet has life. (Credit: Aaron Gronstal)}
\label{fig1}
\end{figure}

\subsection{Novel candidate biosignatures}\ \ There has been a large expansion in the proposed biosignatures for the community to consider. For photosynthetic pigments, organisms that extend the wavelengths of light that can drive oxygenic photosynthesis have been discovered (Ho et al. 2016; Li et al., 2015), increasing the types of star-planet combinations that can sustain this metabolism (Takizawa et al., 2017). Surface pigments other than those used for oxygenic photosynthesis have also been proposed, including bacteriorhodopsin and other pigments (e.g., Schwieterman et al., 2015a, Hegde et al., 2015). For atmospheric biosignatures, several thousand volatile gases have been identified as worthy of further consideration (Seager et al., 2016). On planets lacking oxygen, atmospheric features such as organic hazes have also been identified as possible signs of life (Arney et al., 2016). Sustained efforts at formal cataloguing of the new wealth of biosignature features are critically needed. 

\subsection{3D simulation of living worlds}
Modeling tools have become critical in simulating biosignatures on a global scale. These include photochemical and climate models that can self-consistently simulate these biosignatures within their planetary context. A significant advance in this area since 2015 is the utilization of 3-dimensional (3D) spectral models (e.g., Robinson et al., 2011; Schwieterman et al., 2015b). 3D general circulation models (GCMs) are emerging as important theoretical tools to explore the dynamics of planetary climates and to expand conceptualization of the habitable zone (e.g., Turbet et al.2016; Way et al., 2017). Further development of these modeling capabilities will be needed to apply coupled biosphere-atmosphere processes to simulate biosignatures in a planetary systems science context.

\subsection{The importance of environmental context}
Oxygen-based biosignatures (O\textsubscript{2} and/or O\textsubscript{3}) are extremely promising, as they fulfill the three major requirements of a robust atmospheric biosignature: (1) reliability; (2) survivability; and (3) detectability. However, a number of potential ''false positives'' for O\textsubscript{2}/O\textsubscript{3} biosignatures exist, rendering additional environmental context critical for interpreting oxygen-based biosignatures. For example, information about the host star (spectral type, age, activity level), major planet characteristics (size, orbit, mass), and accessory atmospheric species (H\textsubscript{2}O, CO\textsubscript{2}, CO, CH\textsubscript{4}, N\textsubscript{4}) can all help to diagnose pathological high-O\textsubscript{2}/O\textsubscript{3} cases. Similarly, Earth's atmospheric evolution demonstrates that biogenic gases may remain at undetectable levels despite their production by a surface biosphere (Rugheimer and Kaltenegger, in press). 

Planetary characteristics that may enhance the likelihood of such ''false negatives'' should be considered when selecting targets for biosignature searches. Careful selection of targets can help mitigate against the likelihood of false positive O\textsubscript{2}/O\textsubscript{3} signals. For example, selection of older F, G, K or early M dwarf targets (M0-M3) would help guard against false positive O\textsubscript{2}/O\textsubscript{3} signals associated with water loss, while potentially increasing the probability that biogenic O\textsubscript{2}/O\textsubscript{3} will have accumulated to detectable levels. We suggest an integrated observation strategy for fingerprinting oxygenic photosynthetic biospheres on terrestrial planets with the following major steps: (1) planet detection and preliminary characterization; (2) search for O\textsubscript{2}/O\textsubscript{3} spectral features with high-resolution spectroscopy; (3) further characterization and elimination of potential false positives; (4) detailed characterization and the search for secondary biosignatures. The identification of a pigment spectral feature would be a particularly complementary biosignature O\textsubscript{2}/O\textsubscript{3} detection, because it would be consistent with the hypothesis that the O\textsubscript{2} was generated by oxygenic photosynthesis. To further improve confidence in identifying surface signs of photosynthesis, the reflection spectra of the mineral background must also be characterized. Newly developed measurements such as the linear and circular polarization spectra of chiral biomolecules can potentially help rule out such false positives. In addition, models that address the surface coverage of a planet are needed to better understand the detectability of these signals.

\section{Scientific Progress in the Next 20 Years}
\subsection{Cross-disciplinary quantitative frameworks}
Much of the top-level theory of biosignatures is described in qualitative terms, and the associated advice to mission/instrument design teams is similarly qualitative. For example, we know that the confirmation of biosignatures requires a comprehensive classification of the planetary environment, which in turn leads to a suggestion to obtain observations with as broad of a wavelength range as possible. But evaluation of detailed trade-offs for specific instruments, and eventually the interpretation of data from biosignature searches, will be best enabled by a more quantitative framework.

A major challenge in such quantification is that assessing the presence or absence of life on a planet is an inherently complex problem, requiring comprehensive analyses of the planetary context. And a planet will have multiple systems that interact with each other, often in nonlinear ways. Accounting for this in a quantified manner -- and doing so in a way that is flexible enough to handle alien worlds with potentially alien climates and potentially alien life - requires an encompassing framework. At the EBWWW, a variety of approaches were discussed, including: process-based planet systems models; quantification of thermodynamic and/or kinetic disequilibrium in a planet's atmosphere (after Krissanssen-Totten et al., 2016); assessment of the complexity of atmospheric photochemical networks (after Holme et al. 2011); and utilization of Bayes' Theorem to assess the data from a single planet or a series of planets. Bayes' Theorem, in particular, was identified as having the potential to advance our field's ability to synthesize sparse data, and as a framework for combining understanding from diverse scientific disciplines. 

According to Bayes' Theorem, one can calculate the conditional probability that something is true, such as the likelihood of a system having a given property based on available data.  An example mathematical formalism for exoplanet biosignatures is shown in Figure 2, from Catling et al. (in press). This derivation specifically dissects what might be observed (D = data) given either the presence or absence of life within a particular exoplanet environment context (C = context), i.e., P(data\textbar  context and life) and P(data\textbar context and no life), respectively. The conditional probabilities here account for the intertwining of life with its environment, such that they cannot be independent. P(life\textbar context) is a quantitative expression of likelihood of life given the context of the exoplanet, such as amenability to habitability. This is distinct from P(life) (the probability of life occurring at all in the universe).  The latter might be estimated from how quickly life emerged on Earth, but is truly quantifiable only with large statistics, after more examples of life have been already discovered, which Walker et al. (in review) expand upon.  Bayes' Theorem also provides a means to incorporate uncertainty in data (Parviainen, 2017), additional types and novel concepts of life, such as exotic adaptations, network theory, alternative chemistry, or statistics from ensemble investigations, and in general new data and ideas as they develop (e.g. Deeg et al., 2017). The Bayesian approach thus affords the synthesis of diverse areas of knowledge into a quantitative framework. It also is highly useful for identifying the terms most challenging to quantify. Given the highly interdisciplinary nature of the search for exoplanet biosignatures, adoption of a Bayesian concept is encouraged to help scientists work across disciplines, identify the significance of critical unknowns, and provide quantitative assessments of confidence in scientific conclusions.

\begin{figure}[t]
\centerline{\includegraphics[width=3.5in]{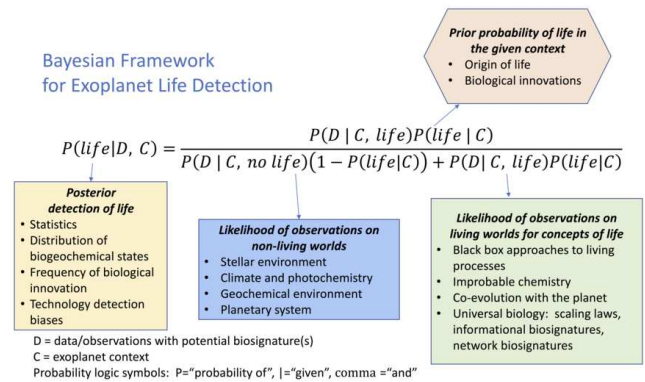}}
\caption{A Bayesian framework, applied to the detection of life on extrasolar planets. Equation from Catling, et al., in press.  Adapted from Walker et al., in review.}
\label{fig2}
\end{figure}

The community is beginning to build comprehensive modeling tools, and the future research directions required to quantify our assessments are reviewed in the EBWWW paper by Walker et al. (in review). The tools for simulating data that could come from inhabited/uninhabited worlds are already under development with both flexible 1-dimensional atmospheric models that can be coupled to subsurface and escape models, and comprehensive but less flexible 3-dimensional global climate models. Current work - by large interdisciplinary teams - is increasing the comprehensiveness of the former models as well as the flexibility of the latter ones. This development of models must continue - and the community involvement in their development must be expanded. We also require advancements in chemistry and biology research on life's origins on Earth, and the environments in which life might originate elsewhere, to help with our assessments of P(life). Finally, we must advance our grasp of the likelihood of certain biological innovations, and better understand the full range of metabolisms life can utilize for obtaining energy, beyond those found on modern-day Earth.

\section{Telescopes in Planning or Development}
The most critical step in our search for extrasolar life is to detect spectroscopic properties of potentially habitable planets. A handful of Earth-sized planets in the HZs of late-type stars have already been identified (Anglada-Escud\'{e} et al., 2016; Dittmann et al., 2017; Gillon et al., 2017), including a few that are close enough for follow-up observation. Soon, discoveries and astrophysical characterization of similar targets will be accelerated by TESS (2018-), CHEOPS (2018-), and ongoing/future ground-based RV surveys. Follow-up observations of such targets could be conducted by the James Webb Space Telescope JWST (2019-), and the next generation ground-based telescopes (GMT, TMT, ELT: 2020s-) and next-generation flagship space telescopes (OST, LUVOIR, HabEx) armed with high-resolution and/or high-contrast instruments. The detectability of the specific features depends on the system properties of the targets as well as the noise floor. And we note that the false positive concerns noted above (as well as concerns about habitability) are greatest for the stellar targets whose planets we will be able to see with this technique. Such concerns should not dissuade us from these observations, but they do make target selection and precursor observations of stellar host properties critical. The characterization of Earth-like HZ planets around Solar-type stars will require more sensitive observations. The PLATO (2026-) mission is specifically targeted at transiting planets in a wider parameter space, including small HZ planets around Solar-type stars. The spectroscopic characterization of potentially Earth-like worlds around Sun-like stars demands space-based high-contrast observations. These observations are not feasible with current and planned facilities, but are among the driving science goals for HabEx and LUVOIR.

\section{Existing and Needed Partnerships}
The EBWWW revealed that the search for exoplanet life is still largely driven by astronomers and planetary scientists, and that this field requires more input from origins of life researchers and biologists to advance a process-based understanding for planetary biosignatures. This includes assessing the prior that a planet may have life, or a life process evolved for a given planet's environment. These advances will require fundamental research into the origins and processes of life, in particular for environments that vary from modern Earth's. Thus, collaboration between origins of life researchers, biologists, and planetary scientists is critical to defining research questions around environmental context. Private partnerships - mostly limited to building spaceflight hardware in the past - must expand to improve our computational and modeling capabilities. These collaborations could include the development of generic research tools, as well as specific collaborations to improve or re-write scientific code. This latter area has tremendous potential for new public-private partnerships, as the codes required to quantify our certainty of a biological detection will be complex, and codes with such complexity should be crafted in partnership with professional programmers.

\section{Realizing NASA's astrobiology goals}
To realize our goals, and to enable probabilistic assessments of whether or not a planet has life, we require the following developments:
\begin{itemize}
 \item A more complete incorporation of biological understanding into the field
  \item Models of fundamental abiotic processes under planetary conditions different than our own
 \item Evaluation of the wealth of potential new biosignatures, both surface and gaseous, and consideration of their false positives
 \item An improved capability to predict the expression of photosynthesis in different stellar-planetary environments
 \item Sustained institutional support to characterize the physical and chemical properties of biogenic small volatile gases
 \item Development and infrastructure support for 3-D general circulation models (GCMs) for exoplanets, to simulate biosignatures in 3-D
 \item Expansion of coupling of 1D planetary models for mantle, atmospheric chemistry, climate, ocean, biology, and atmospheric escape processes, with different stellar inputs, to simulate biosignatures in a planet systems context
 \item More accounting of model uncertainties
 \item Finally, a Bayesian framework to foster integration of diverse scientific disciplines and to accommodate new data and novel concepts is advocated for further development in the classroom and in collaborative research
\end{itemize}

That last goal is critical, as a quantitative approach will advance our field in multiple ways. For exoplanet astrobiologists, it will be a powerful way to consider future mission/instrument trade-offs, or to inform future target selection. For our astrobiology peers searching for life on planets around other stars, it will provide a comparative tool with different proposed biosignatures for other targets. For our scientific colleagues beyond astrobiology, it will provide a rigorous test of our conclusions. And for the general public and to stakeholders, it will lead to the ability to clearly and consistently communicate our level of confidence that we are not alone.

\section{References}
Anglada-Escud\'{e}, G., et al. (2016) Nature 536:437. 

Arney G., et al. (2016) Astrobiology 16:873.

Catling D.C., et al. (in press) Astrobiology. arXiv:1705.06381


Deeg H. J., et al. (2017) Handbook of Exoplanets, Springer.

Des Marais D.J., et al. (2002) Astrobiology 2:153.

Dittmann, J.A., et al. (2017) Nature 544:333.

Fujii Y., et al. (in review) Astrobiology. arXiv:1705.07098

Gillon M., et al. (2017) Nature 542:456.

Hegde S., et al. (2015) PNAS 112:3886. 

Ho M.Y., et al. (2016) Science 353:9178.

Holme P., et al. (2011) PLoS ONE 6:19759.

Krissanssen-Totton J., et al. (2016) Astrobiology, 16:39.

Li Y.Q., et al. (2015) Funct. Plant Biol. 42:493.

Luger R., et al. (2015) Astrobiology 15:119.

Meadows V.S., et al. (in press) Astrobiology. arXiv:1705.07560


Parviainen H. (2017)  In: Handbook of Exoplanets, Springer pp 1-24.

Robinson, T.D., et al. (2011) Astrobiology 11:393.

Rugheimer and Keltenegger (in press) Astrophys. Journal. arXiv:1712.10027


Schwieterman E.W., et al. (2015a) Astrobiology 15:341.

Schwieterman, E.W., et al. (2015b) Astrophys. Journal 810:57. 

Schwieterman E.W., et al. (in press) Astrobiology. arXiv:1705.05791

Seager S., et al. (2016) Astrobiology 16:465.

Takizawa K., et al. (2017) Nature Sci. Reports 7:id.7561

Turbet M., et al. (2016) Astron. Astrophys. 596.

Walker S.I., et al. (in review) Astrobiology. arXiv:1705.08071

Way M.J., et al. (2017) Astrophys. Journal Suppl. Series 231:12.

Wordsworth R., et al. (2014) Astrophys. Journal Letters, 785.

\end{article}









\end{document}